\documentstyle[11pt,epsf,paspconf]{article}

\begin{document}

\title{WOMBAT \& FORECAST:  \\Making Realistic Maps of the Microwave Sky}

\author{Andrew H. Jaffe\altaffilmark{1}}
\affil{Center for Particle Astrophysics, 301 LeConte, University of California,
    Berkeley, CA 94720}

\author{
Eric Gawiser,
Douglas Finkbeiner,
Joanne C. Baker,
Amedeo Balbi,
Marc Davis,
Shaul Hanany,
William Holzapfel,
Mark Krumholz,
Leonidas Moustakas\altaffilmark{2},
James Robinson,
Evan Scannapieco,
George F. Smoot,
\& Joseph Silk\altaffilmark{3}
}
\affil{Center for Particle Astrophysics, Departments of Astronomy and
  Physics, University of California, Berkeley 94720}
\altaffiltext{1}{jaffe@cfpa.berkeley.edu} 
\altaffiltext{2}{Also, Department of Astrophysics, University of Oxford}
\altaffiltext{3}{Also, Department of Physics, University of Oxford, Nuclear
  and Astrophysics Laboratory, Keble Road, Oxford OX1 3RH.}
\begin{abstract}
  The Wavelength-Oriented Microwave Background Analysis Team (WOMBAT) is
  constructing microwave maps which will be more realistic than previous
  simulations.  Our foreground models represent a considerable
  improvement: where spatial templates are available for a given
  foreground, we predict the flux and spectral index of that component
  at each place on the sky and estimate uncertainties.  We will produce
  maps containing simulated CMB anisotropy combined with expected
  foregrounds.  The simulated maps will be provided to the
  community as the WOMBAT Challenge, so such
  maps can be analyzed to extract cosmological parameters by scientists
  who are unaware of their input values.  This will test the efficacy of
  foreground subtraction, power spectrum analysis, and parameter
  estimation techniques and help identify the areas most in need of
  progress.  These maps are also part of the FORECAST project, which
  allows web-based access to the known foreground maps for the planning
  of CMB missions.
\end{abstract}
\section{Introduction}
Cosmic Microwave Background (CMB) anisotropy observations during the
next decade will yield data of unprecedented quality and quantity.
Determination of cosmological parameters to the precision that has been
forecast (Jungman et al.\ 1996, Bond, Efstathiou, \& Tegmark 1997,
Zaldarriaga, Spergel, \& Seljak 1997, Eisenstein, Hu, \& Tegmark 1998)
will require significant advances in analysis techniques to handle the
large volume of data, subtract foregrounds, and account for systematics.
We must ensure that these techniques do not introduce biases into the
estimation of cosmological parameters.

The Wavelength-Oriented Microwave Background Analysis Team (WOMBAT,
http://astro.berkeley.edu/wombat, see also Gawiser et al 1998) will
produce state-of-the-art foreground simulations, using all available
information about frequency and spatial dependence.  Phase information
(detailed spatial morphology) offers the possibility of improving upon
techniques that only use the angular power spectrum of the
foregrounds to account for their distribution.  Most techniques assume
the frequency spectra of the components is constant across the sky, but
we will provide information on the spatial variation of each component's
spectral index whenever possible.  This reflects our actual sky; with
the high precision expected from future CMB maps we must test our
techniques on as realistic a map as possible.  A second advantage is the
construction of a common, comprehensive database for all known CMB
foregrounds, {\em including
  uncertainties}. 

These models provide the starting point for the WOMBAT Challenge, in
which we will generate maps for various cosmological models and offer
them to the community for analysis without revealing the input
parameters.
The WOMBAT Challenge promises to shed light on several open questions in
CMB data analysis: What are the best foreground subtraction techniques?
Will they allow instruments such as MAP and Planck to achieve the
precision in $C_\ell$ reconstruction which has been advertised, or will
errors increase significantly due to foreground uncertainties?  Perhaps
most importantly, do some CMB analysis methods produce biased estimates
of the cosmological parameters?

\section{Microwave Foregrounds} 


There are four major expected sources of Galactic emission at microwave
frequencies: thermal emission from dust, electric or magnetic dipole
emission from spinning dust grains (Draine \& Lazarian 1998a,1998b),
free-free emission from ionized hydrogen, and synchrotron radiation from
electrons accelerated by the Galactic magnetic field.  Good spatial
templates exist for thermal dust emission (Schlegel, Finkbeiner, \&
Davis 1998 [SFD]) and synchrotron emission (Haslam et al.\  1982),
although the $0\fdg5$ resolution of the Haslam maps means that
smaller-scale structure must be simulated.  Extrapolation to microwave
frequencies is possible using maps which account for spatial variation
of the spectra (Finkbeiner, Schlegel, \& Davis 1999; Platania et al.\ 
1998).

A spatial template for free-free emission based on observations of
H$\alpha$ (Smoot 1998, Marcelin et al.\ 1998) can be created in the near
future by combining WHAM observations (Haffner, Reynolds, \& Tufte 1998)
with the southern celestial hemisphere H$\alpha$ Sky Survey (McCullough
1998).  While it is known that there is an anomalous component of
Galactic emission at 15-40 GHz (Kogut et al.\ 1996, Leitch et al.\ 1997,
de Oliveira-Costa et al.\ 1997) partially correlated with dust
morphology, it is not yet clear whether this is spinning dust grain
emission or free-free emission somehow uncorrelated with H$\alpha$
observations.  In fact, spinning dust emission {\em per se} has yet to
be observed, so uncertainties in its amplitude are tremendous.  A
template for this ``anomalous'' component will have large uncertainties.

Three nearly separate categories of galaxies will also generate
foreground emission: radio-bright galaxies, low-redshift
IR-bright galaxies, and high-redshift IR-bright galaxies.  The
anisotropy from these foregrounds is predicted by Toffolatti et
al.\ (1998) using models of galaxy evolution to produce source counts,
and updated models calibrated to recent SCUBA observations are 
available (Blain, Ivison, Smail, \& Kneib 1998, Scott \& White 1998).
For the high-redshift SCUBA galaxies, no spatial template is
available, so a simulation with realistic clustering will be necessary.
Scott \& White (1998) and Toffolatti et al.\ (1998) have used very
different estimates of clustering,
so this issue will need to be looked at more carefully.  
Limits on anisotropy generated by high-redshift galaxies
and as-yet-undiscovered types of point sources are given by Gawiser,
Jaffe, \& Silk (1998) using recent observations over a wide range of
frequencies.  Their upper limit of $\Delta T/T=10^{-5}$ for a
$10'$ beam at 100 GHz is a sobering result.
The 5319
brightest low-redshift IR galaxies detected at 60$\mu$m are in
the IRAS 1.2 Jy catalog (Fisher et al.\ 1995) and can be extrapolated to
100 GHz with a systematic uncertainty of a factor of a few (Gawiser \&
Smoot 1997). 
Sokasian, Gawiser, \& Smoot (1998) have compiled a catalog of
2200 bright radio sources, some of which have been observed at 90 GHz
and fewer still above 200 GHz.  They have
developed a method to extrapolate spectra with a
factor of two uncertainty at 90 GHz.

Secondary CMB anisotropy is generated as CMB photons are scattered after the
original last-scattering surface.  The most important of these effects
occurs as the shape of the blackbody spectrum is altered through inverse
Compton scattering by the thermal Sunyaev-Zel'dovich (1972; SZ) effect.
Simulations have been made of the impact of SZ in large-scale structure
(Persi et al.\ 1995), clusters (Aghanim et al.\  1997) and groups (Bond \&
Myers 1996).
The brightest 200 X-ray clusters known from the
XBACS catalog can be used to incorporate the locations of the
strongest SZ sources (Refregier, Spergel, \& Herbig 1998).  

\begin{figure*}[htbp]
  \begin{center}
    \leavevmode\epsfxsize=2.4in\epsfbox{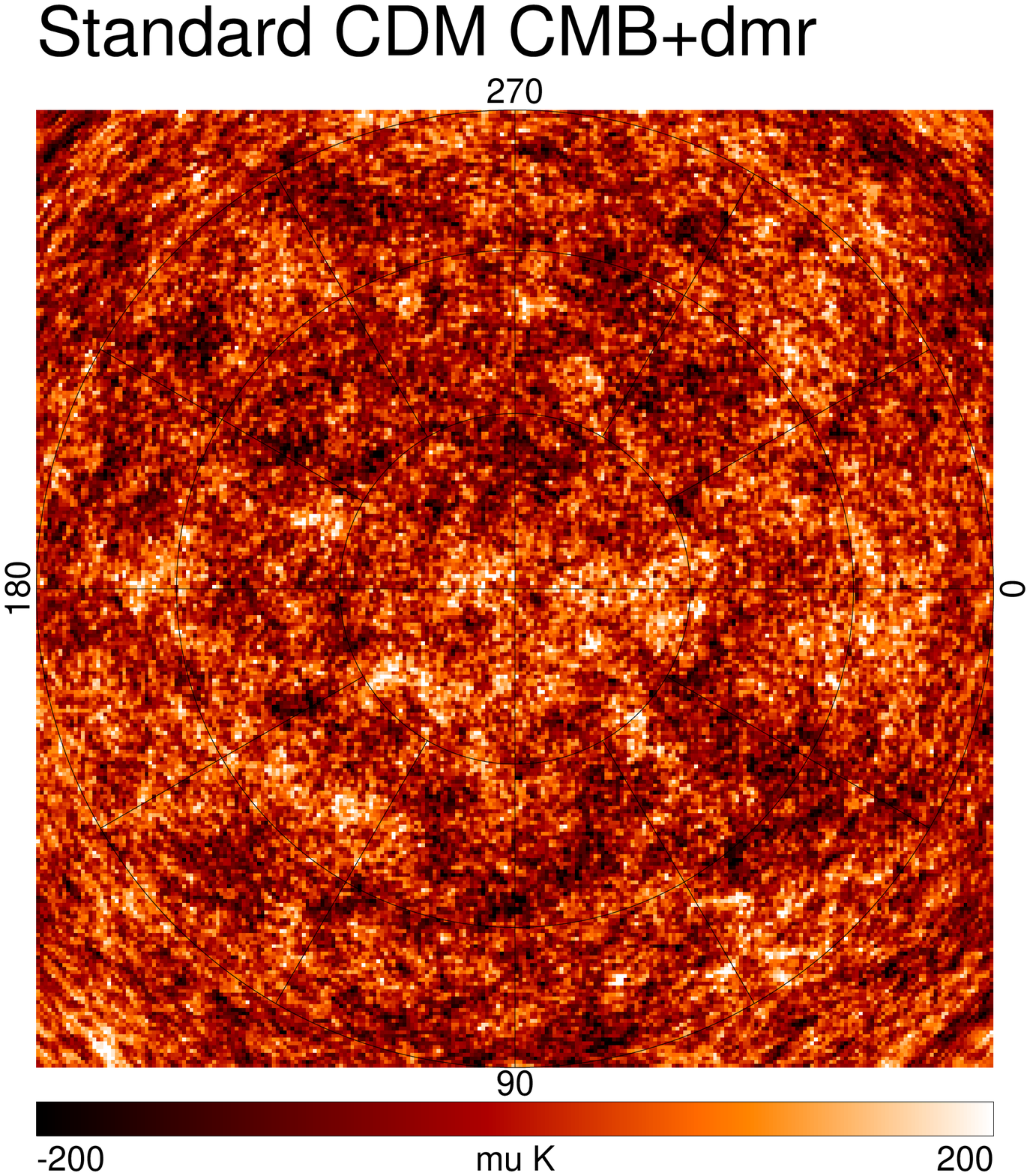}
    \epsfxsize=2.4in\epsfbox{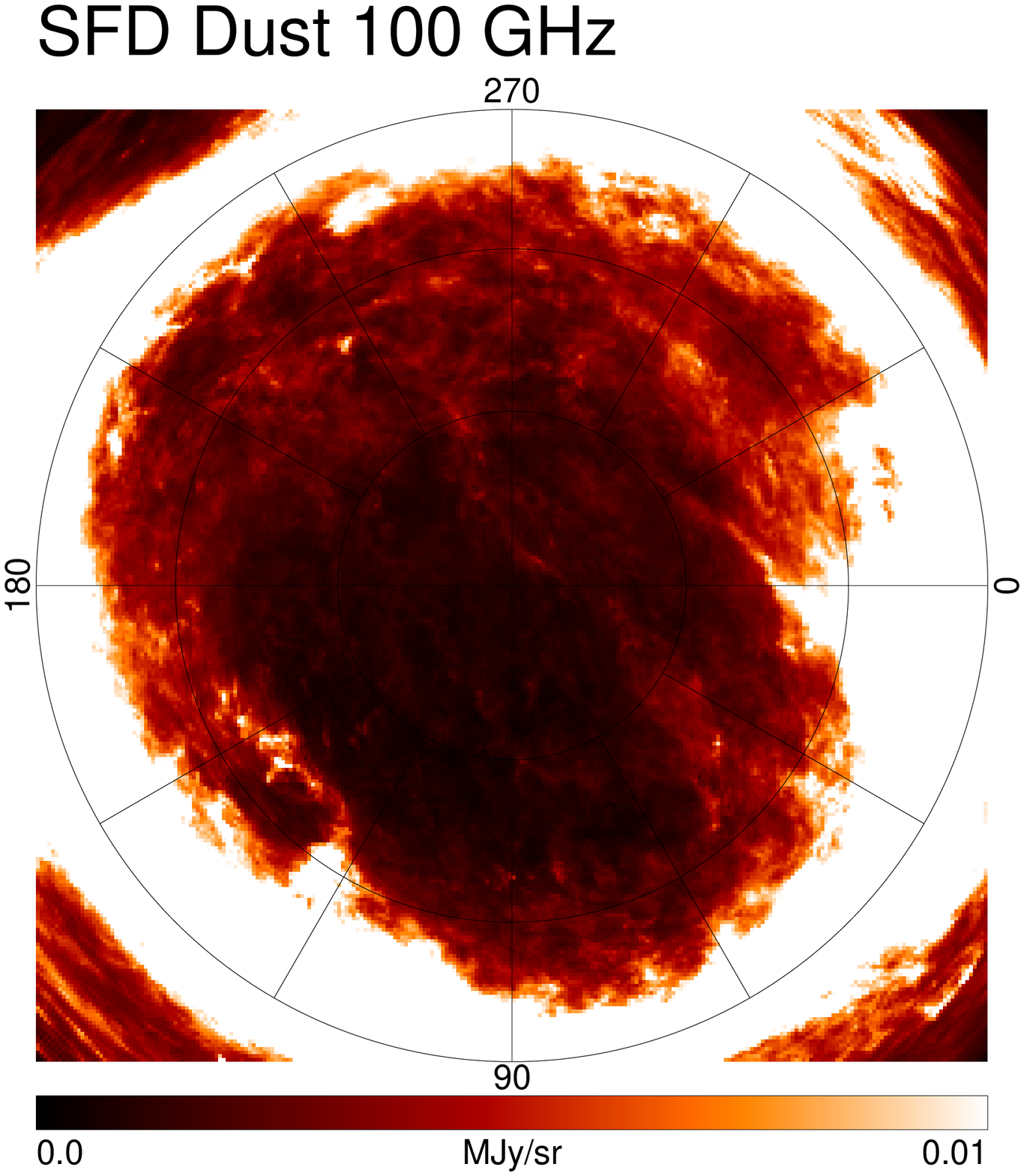}
  \end{center}
  \begin{center}
    \leavevmode\epsfxsize=2.4in\epsfbox{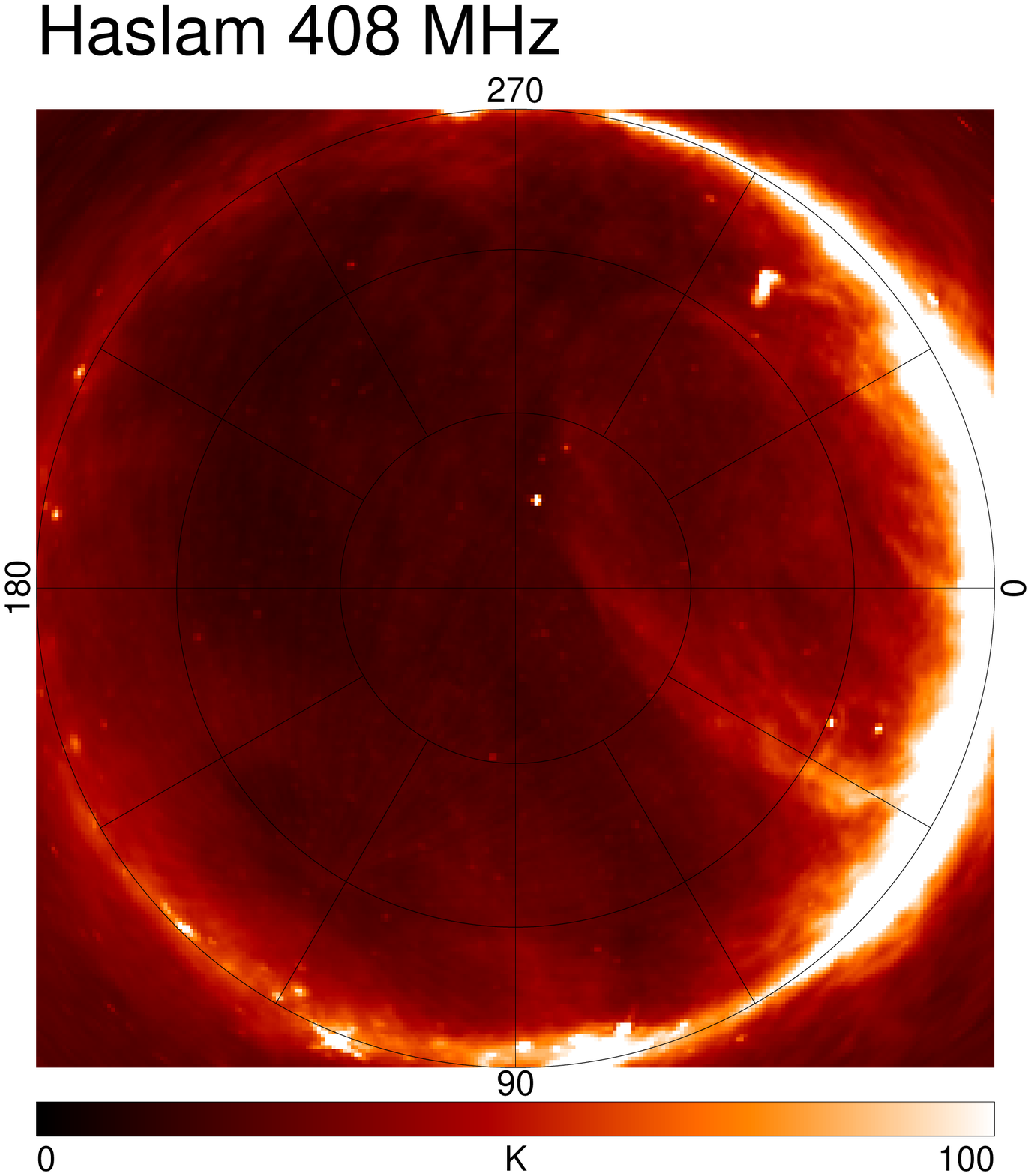}
    \epsfxsize=2.4in\epsfbox{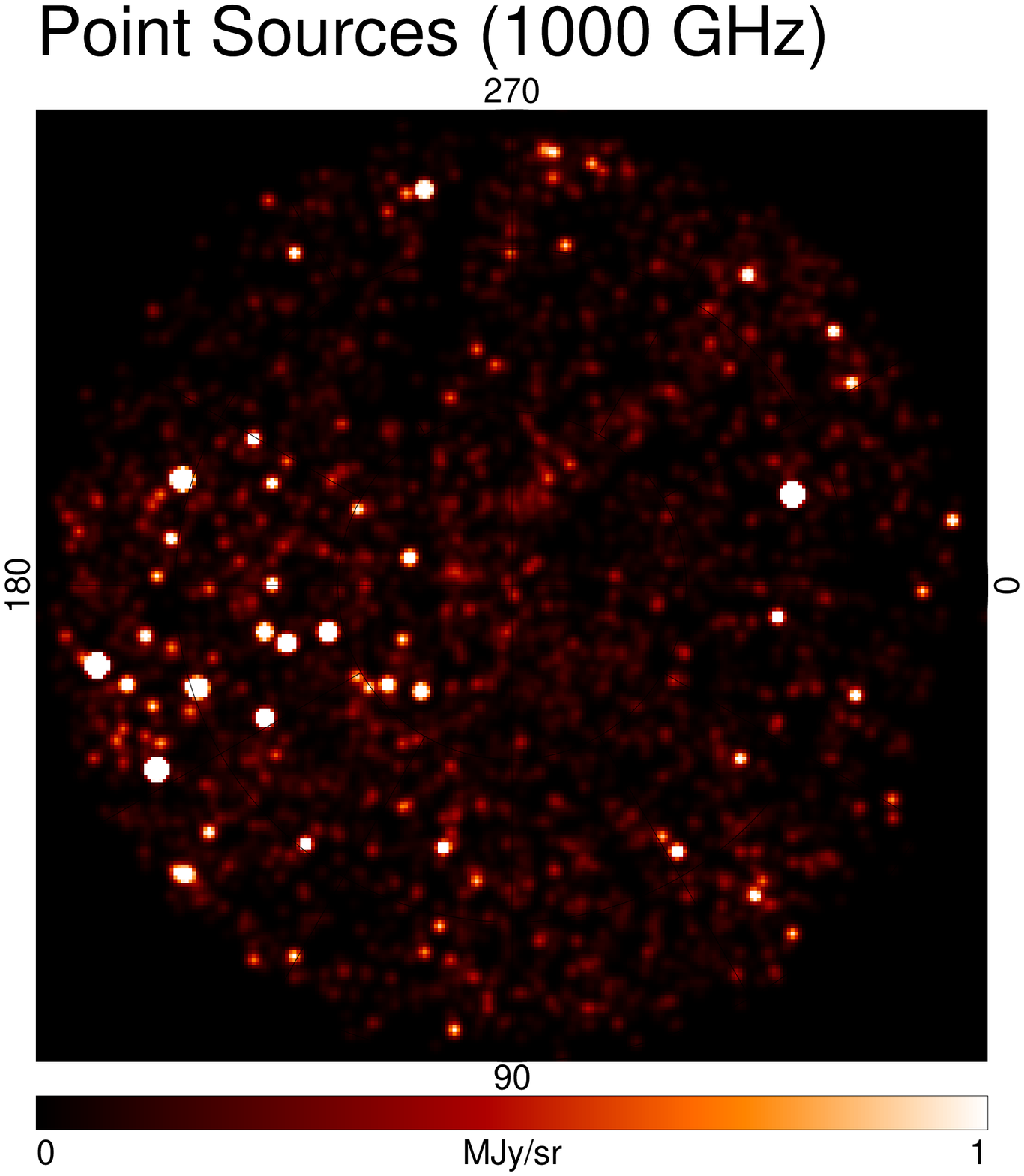}
  \end{center}
  \caption{Foreground maps of the Northern Galactic Hemisphere, as 
    labeled. The first author (AHJ) apologizes for the poor resolution
    and color scale.
\label{fig:wombatinput}}
\end{figure*}

In Figure~\ref{fig:wombatinput}, we show an example of some of the
foreground maps we will use: the CMB itself (a realization of standard
CDM constrained to the COBE/DMR results, courtesy of E.~Scannapieco),
the SFD dust map, the Haslam synchrotron map, and the IR and radio
source catalog amassed by Gawiser et al.\ Outside of the galactic plane,
the morphology of each component is quite distinct.

\section{Reducing Foreground Contamination}

Various methods have been proposed for reducing foreground
contamination.  For point sources, it is possible to mask pixels which
represent positive $5 \sigma$ fluctuations since these are highly
unlikely for Gaussian-distributed CMB anisotropy and can be assumed to
be caused by point sources.  This technique can be improved somewhat by
filtering (Tegmark \& de Oliveira-Costa 1998; see Tenorio et al.\ 1998
for a different technique using wavelets).  Sokasian, Gawiser, \& Smoot
(1998) demonstrate that using prior information from good catalogs may
allow the masking of pixels which contain sources brighter than $1
\sigma$.  For the 90 GHz MAP channel, this could reduce the residual
radio source contamination by a factor of two.  Galactic foregrounds with
well-understood spectra can be projected out of multi-frequency
observations on a pixel-by-pixel basis (Dodelson \& Kosowsky 1995,
Brandt et al.\ 1994).

The methods for foreground subtraction which have the greatest level of
sophistication and have been tested most thoroughly ignore the known
locations on the sky of some foreground components.  Multi-frequency
Wiener filtering uses assumptions about the spatial power spectra and
frequency spectra of the components to perform a separation in spherical
harmonic or Fourier space (Tegmark \& Efstathiou 1996; Bouchet et al.\ 
1995,1997,1998; Knox 1998).  However, it does not include any phase
information.  The MaxEnt Method (Hobson et al.\ 1998a) can add
phase information on diffuse Galactic foregrounds in small patches of
sky but treats extragalactic point sources as an additional source of
instrument noise, with good results for simulated Planck data (Hobson et
al.\ 1998b) and worrisome systematic difficulties for simulated MAP data
(Jones, Hobson, \& Lasenby 1998).  
Both methods have difficulty if pixels are masked
due to strong point source contamination or the spectral indices of the
foregrounds are not well known (Tegmark 1998).

Since residual contamination can increase uncertainties and bias
parameter estimation, it is important to reduce it as much as possible.
Current analysis methods usually rely on cross-correlating the CMB maps
with foreground templates at other frequencies (see de Oliveira-Costa et
al.\ 1998; Jaffe, Finkbeiner, \& Bond 1999).  It is clearly superior to
have localized information on extrapolation of these templates to the
observed frequencies; otherwise this cross-correlation only identifies
the emission-weighted average spectral index of the foreground.

When a known foreground template is subtracted from a CMB map, it is
inevitable that the amplitude used
will be slightly different from the true value.  This leads to
off-diagonal structure in the ``noise'' covariance matrix of the
remaining CMB map, as opposed to the contributions of expected CMB
anisotropies which gives diagonal contributions to the covariance matrix
of the $a_{\ell m}$.  Thus incomplete foreground subtraction, like $1/f$
noise, can introduce correlations into the covariance matrix of the
$a_{\ell m}$.  These complicate the likelihood analysis necessary for
parameter estimation (Knox 1998), but phase information should
reduce inaccuracies in foreground subtraction.

\section{The WOMBAT Challenge}

Our purpose in conducting a ``hounds and hares'' exercise is to simulate
the process of analyzing microwave maps as accurately as possible.
We will make our knowledge of the various foreground components
available, and each best-fit foreground map will be accompanied by its
uncertainties and possible systematic errors.  Each simulation of a
foreground will incorporate a realization of those uncertainties.  Very
little is known about the locations of high-redshift IR-bright galaxies
and SZ-bright clusters, so WOMBAT will provide simulations of these
components.  The rough characteristics of these high-redshift sources,
but not their locations, will be revealed.  This simulates the real
observing process in a way not achieved by previous work.

One of the biggest challenges in real-world observations is being
prepared for surprises, both instrumental and astrophysical (see Scott
1998 for an eloquent discussion); we will include a few in our maps.

We will release our maps for the community to subtract the foregrounds
and extract cosmological information.  The WOMBAT Challenge is scheduled
to begin on March 15, 1999 and will offer participating groups four
months to analyze the maps and report their results.\footnote{see
  http://astro.berkeley.edu/wombat for timeline, details for
  participants, and updates} We will produce simulations analogous to
high-resolution balloon observations (e.g. MAXIMA and BOOMERANG; see
Hanany et al.\ 1998 and de Bernardis \& Masi 1998) and the MAP satellite
(10$^6$ pixels at 13$'$ resolution for a full-sky
map)\footnote{http://map.gsfc.nasa.gov}.
We plan to use the HEALPIX package of pixelization and analysis
routines\footnote{http://www.tac.dk/\~{}healpix}.  We provided a
calibration map of CMB anisotropy with a disclosed angular power
spectrum in January 1999 so that participants could test the download
procedure and become familiar with HEALPIX.  Groups who 
participate 
will be asked to provide us with a summary of their
analysis techniques.  They may choose to remain anonymous in our
comparison of the results but are encouraged to publish their own
conclusions.
\begin{figure*}[htbp]
  \begin{center}
    \leavevmode\epsfxsize=2.4in\epsfbox{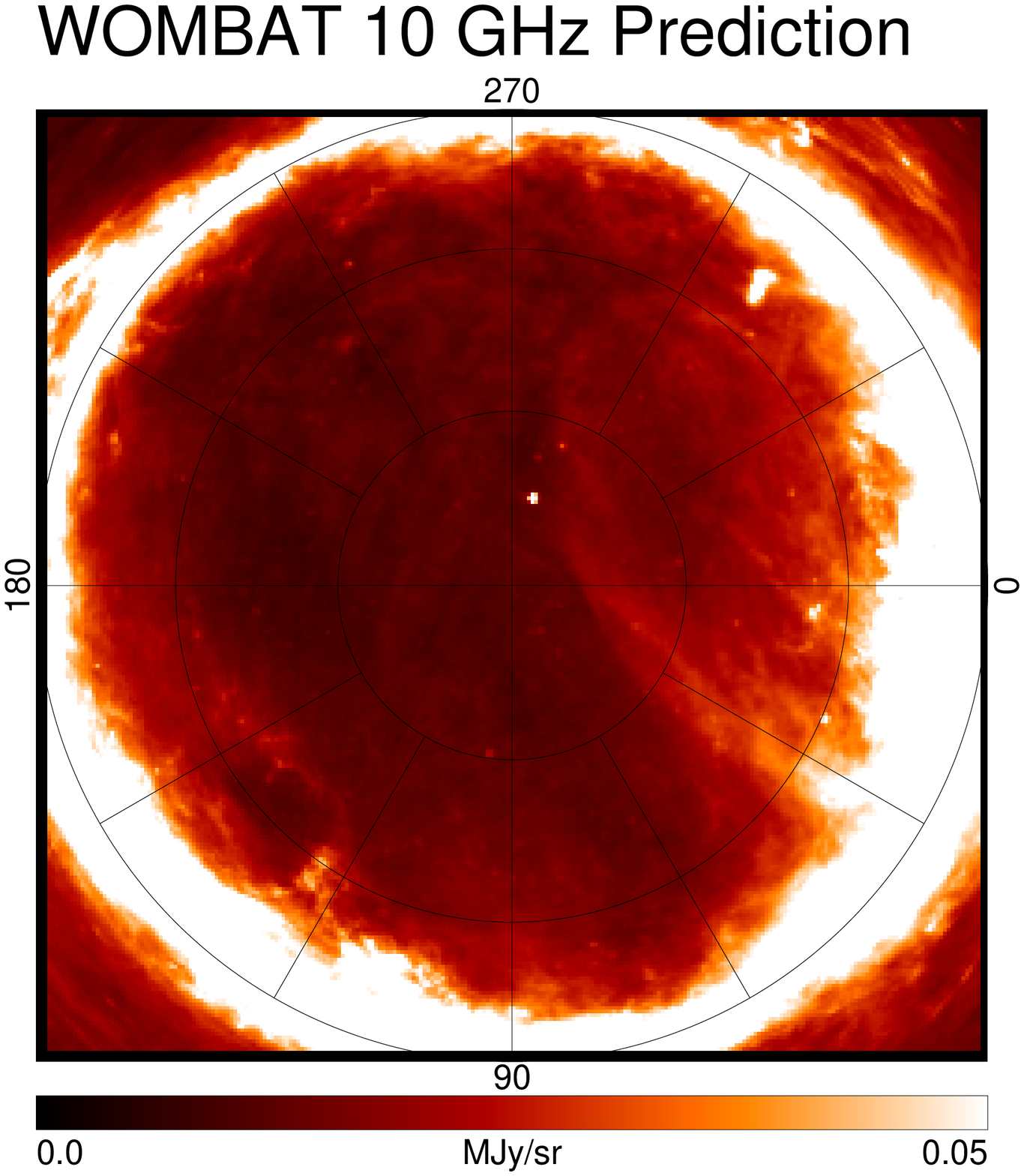}
    \epsfxsize=2.4in\epsfbox{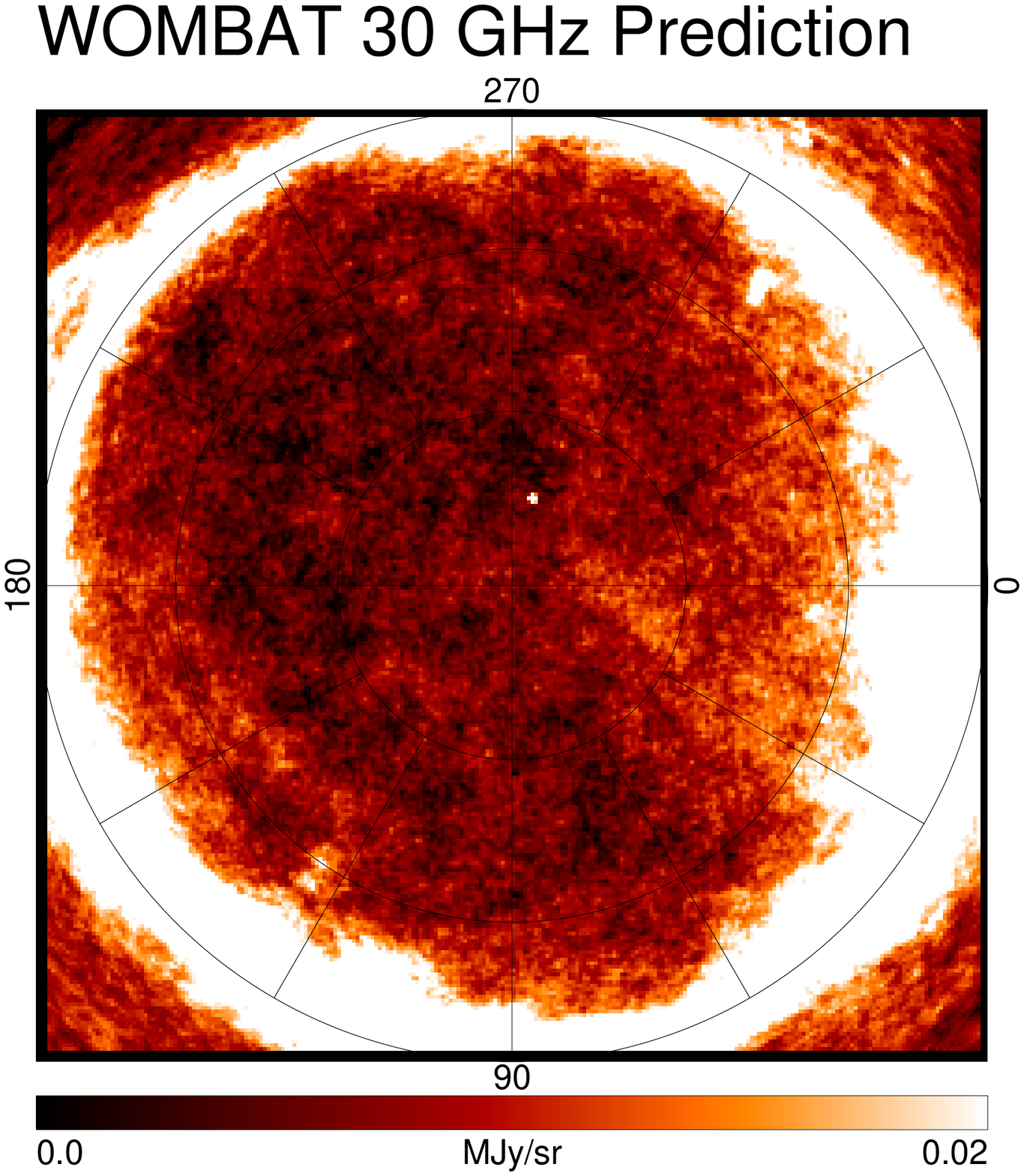}
  \end{center}
  \begin{center}
    \leavevmode\epsfxsize=2.4in\epsfbox{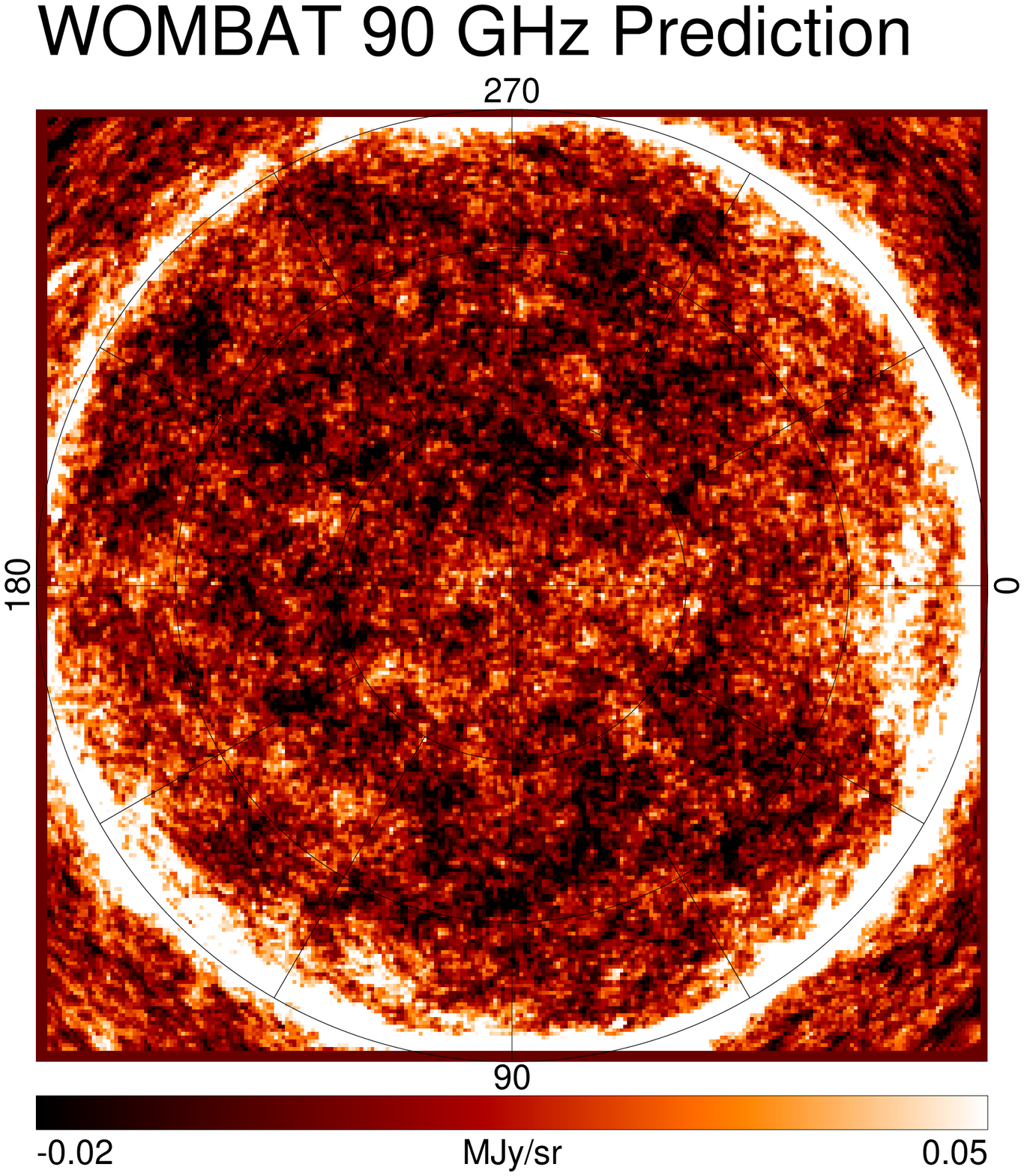}
    \epsfxsize=2.4in\epsfbox{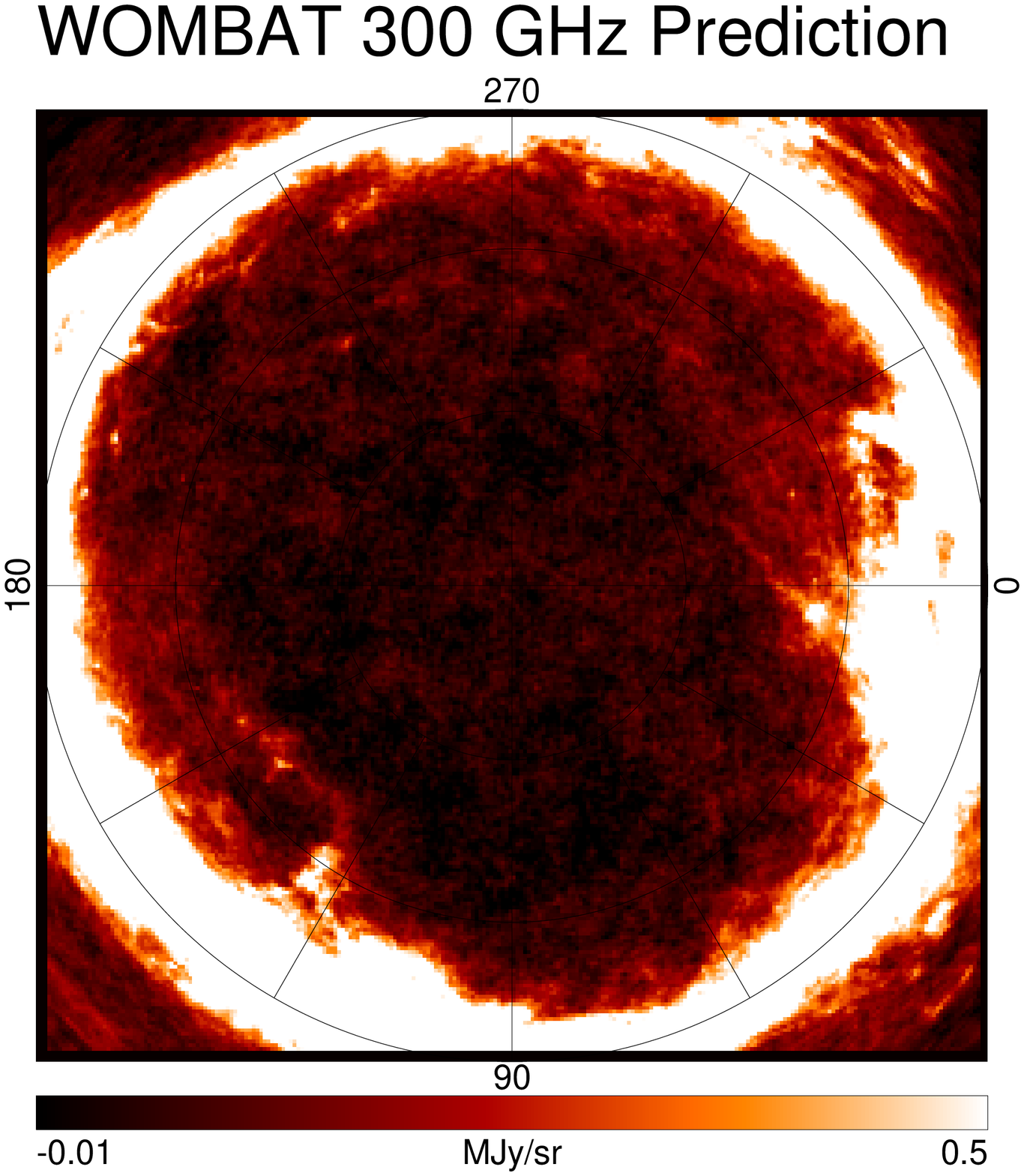}
  \end{center}
  \caption{WOMBAT maps of the Northern Galactic hemisphere, extrapolating the
    maps of figure 1 to the frequencies labeled.  The first author (AHJ)
    apologizes for the poor resolution and color
    scale.\label{fig:wombatoutput}}
\end{figure*}
In Figure~\ref{fig:wombatoutput}, we show a very simple example of what
we will produce. It extrapolates the maps and catalogs of
Figure~\ref{fig:wombatinput} to frequencies of 10--300~GHz. At low
freqeuencies, the maps (away from the galactic plane) are dominated by
synchrotron emission, at 90~GHz by the CMB itself, and at 300~GHz by
dust (and by extragalactic point sources which are not easily visible at
this resolution). Visually, some sort of separation of the components
seems simple, but doing it at the high precision necessary (and claimed)
for CMB parameter determination to ``unprecedented accuracy'' remains a
challenge. 


\section{FORECAST}


The other thrust of the microwave mapmaking effort is to aid in the
planning of future CMB anisotropy missions.  We will enable quick and
easy access to the foreground maps, combined with our best-guess
extrapolations to experimental frequencies.  Because uncertain
extrapolation is involved, we will also provide {\it errors} on the
results. Given specific information about the observing strategy,
observers will be able to quickly call up predictions for their
experiment's contamination by foreground emission.




\section{Conclusions}

Undoubtedly the most important scientific contribution that WOMBAT will
make is the production of realistic full-sky maps of all major microwave
foreground components with estimated uncertainties.  These maps are
needed for foreground subtraction and estimation of residual foreground
contamination in present and future CMB anisotropy observations.  With
FORECAST, instrumental teams will be able to conduct realistic
simulations without needing to assume overly idealized models for the
foregrounds.  By combining various realizations of these foreground maps
within the stated uncertainties with a simulation of the intrinsic CMB
anisotropies, we will produce the best simulations
so far of the microwave sky.  

We can test the resilience of CMB
analysis methods to surprises such as unexpected foreground amplitude or
spectral behavior, correlated instrument noise, and CMB fluctuations
from non-gaussian or non-inflationary models.  Cosmologists need to know
if such surprises can lead to the misinterpretation of cosmological
parameters. 

Perhaps the greatest advance we offer is the ability to evaluate the 
importance of studying the detailed locations of foreground sources.  
It may turn out that techniques which use phase information are needed
in order to reduce foreground contamination to a level which does not
seriously bias the estimation of cosmological parameters.  Combining
various techniques may lead to improved foreground subtraction methods,
and we hope that a wide variety will be tested by the participants in
the WOMBAT Challenge.

\acknowledgments

We thank Rob Crittenden (IGLOO) and Kris Gorski, Eric Hivon, and Ben
Wandelt (HEALPIX) for making pixelization schemes available to the
community.  We appreciate helpful conversations with Nabila Aghanim,
Giancarlo de Gasperis, Alex Refregier, David Schlegel, and Philip Stark.  
Some of the
work described here is done under the auspices of the COMBAT
collaboration supported by NASA AISRP grant NAG-3941 and NASA LTSA grant
NAG5-6552.

\end{document}